\documentclass[sigconf]{acmart}

\copyrightyear{2025}
\acmYear{2025}
\setcopyright{acmlicensed}\acmConference[WWW Companion '25]{Companion Proceedings of the ACM Web Conference 2025}{April 28-May 2, 2025}{Sydney, NSW, Australia}
\acmBooktitle{Companion Proceedings of the ACM Web Conference 2025 (WWW Companion '25), April 28-May 2, 2025, Sydney, NSW, Australia}
\acmDOI{10.1145/3701716.3717740}
\acmISBN{979-8-4007-1331-6/2025/04}

\AtBeginDocument{%
  }

\usepackage{xspace}
\usepackage{booktabs}
\usepackage{todonotes}
\usepackage{tabularx}
\usepackage{subcaption}
\usepackage{colortbl} 
\usepackage{multirow}
\usepackage{color,soul}
\usepackage[normalem]{ulem}

\usepackage{multicol}
\usepackage{float}

\usepackage{tabularray}
\usepackage{float}
\usepackage{graphicx}
\usepackage{codehigh}
\usepackage[normalem]{ulem}
\UseTblrLibrary{booktabs}
\UseTblrLibrary{siunitx}

\NewTableCommand{\tinytableDefineColor}[3]{\definecolor{#1}{#2}{#3}}

\newcommand{\one}{({\em i}\/)\xspace}
\newcommand{\two}{({\em ii}\/)\xspace}
\newcommand{\three}{({\em iii}\/)\xspace}

\newcommand{\age}{{\textit{Age}}\xspace}
\newcommand{\gender}{{\textit{Gender}}\xspace}
\newcommand{\health}{{\textit{Health}}\xspace}
\newcommand{\ethnicity}{{\textit{Ethnicity}}\xspace}
\newcommand{\sexualOrientation}{{\textit{Sexual Orientation}}\xspace}
\newcommand{\religion}{{\textit{Religion}}\xspace}
\newcommand{\job}{{\textit{Job}}\xspace}

\newcommand{\physical}{{\textit{Physical Appearance}}\xspace}
\newcommand{\relationship}{{\textit{Relationship}}\xspace}

\def\eg{\emph{e.g.}\xspace}

\def\ie{\emph{i.e. }\xspace}

\newcommand{\pb}[1]{\vspace{0.75ex}\noindent{\bf \em #1}\hspace*{.3em}}

\newcommand\ehsan[1]{\textcolor{black}{#1}}	

\begin{document}


\title[Exploring Self-Disclosure Norms, Engagement Dynamics, and Privacy Implications]{Unpacking the Layers: Exploring Self-Disclosure Norms, Engagement Dynamics, and Privacy Implications}


 \author{Ehsan-Ul Haq}
 \email{euhaq@hkust-gz.edu.cn}
 \orcid{1234-5678-9012}
\affiliation{%
 \institution{Hong Kong University of Science and Technology (GZ)}
 \country{China}
 }

  \author{Shalini Jangra}
  \email{ s.jangra@surrey.ac.uk}
 \orcid{0000-0002-0265-6770}
\affiliation{%
 \institution{University of Surrey}
 \country{UK}
 }

\author{Suparna De}
  \email{ s.de@surrey.ac.uk}
 \orcid{0000-0001-7439-6077}
\affiliation{%
 \institution{University of Surrey}
 \country{UK}
 }

   \author{Nishanth Sastry}
  \email{ n.sastry@surrey.ac.uk}
 \orcid{0000-0002-4053-0386}
\affiliation{%
 \institution{University of Surrey}
 \country{UK}
 }

 \author{Gareth Tyson}
  \email{gtyson@ust.hk}
 \orcid{0000-0003-3010-791X}
\affiliation{%
 \institution{Hong Kong University of Science and Technology (GZ)}
 \country{China}
 }

\renewcommand{\shortauthors}{Haq et. al.}

\begin{abstract}
This paper characterizes the self-disclosure behavior of Reddit users across 11 different types of self-disclosure. We find that at least half of the users share some type of disclosure in at least 10\% of their posts, with half of these posts having more than one type of disclosure.
We show that different types of self-disclosure are likely to receive varying levels of engagement. For instance, a \sexualOrientation disclosure garners more comments than other self-disclosures. We also explore confounding factors that affect future self-disclosure. We show that users who receive interactions from (self-disclosure) specific subreddit members are more likely to disclose in the future. We also show that privacy risks due to self-disclosure extend beyond Reddit users themselves to include their close contacts, such as family and friends, as their information is also revealed. We develop a browser plugin for end-users to flag self-disclosure in their content. 
\end{abstract}

\maketitle

\section{Introduction} \label{sec:introduction}

Online self-disclosure is sharing information about oneself with other users in online communities ~\cite{wang_modeling_2016,joinson_oxford_2007}. This may include information on health, gender, sexual orientation, or general opinions, \eg liking or disliking a person~\cite{lee_online_2023}. Reasons for disclosure are diverse, including the perceived sense of anonymity \cite{ma_anonymity_2016}, adherence to community norms \cite{chen_i_2024}, and to improve engagement with community members~\cite{chen_revisiting_2018}. There are many benefits of self-disclosure (such as support seeking~\cite{choudhury_mental_2014}). However, it can also raise privacy concerns. 
For example, self-disclosure can lead to serious risks related to mental health~\cite{wood_student_2014,tay_mental_2018,nobels_just_2023} 
and intimacy~\cite{lee_effects_2019}.
Indeed, there have been numerous cases where accidental self-disclosure has led to physical harms, \eg loss of employment~\cite{ott_reputation_2013},  harassment~\cite{lauckner_catfishing_2019}, and geo-tagging~\cite{harrigian_geocoding_2018}.
Consequently, we argue that it is vital to better understand how such privacy-invasive disclosures occur, and develop tools to mitigate such risks.

Although there have been prior works that study self-disclosure online, these are either specific to a particular type of self-disclosure (\eg gender~\cite{mejova_gender_2023}) or specific to a particular type of community, \eg support seeking subreddits~\cite{chen_i_2024} and mental health forums~\cite{choudhury_mental_2014}.
Hence, we know little about the true scale of self-disclosure across diverse disclosure categories and communities.
Key questions include:
\one~How often do users self-disclose, and what types of self-disclosure are made together? 
\two~How common are high-risk self-disclosures? 
\three~What is the effect of self-disclosure specific communities on other users' self-disclosure propensity? Answering such questions is critical for formulating better user support against associated privacy risks.
Yet, the subtle complexities of online self-disclosure raise a number of challenges that must be overcome to study this. 
Specifically, to date, we lack a methodical way of identifying and classifying forms of self-disclosure in online posts. This is further complicated by the fact that self-disclosures are not always atomic. For instance, a user may disclose more than one type of identifiable information within one post or across multiple (seemingly unrelated) posts. This is particularly common on platforms like Reddit, where users may participate in multiple subreddits~\cite{crowley_expressive_2014}. 
Take the following posts (paraphrased to retain anonymity) as an example: \textit{``I am a 20 years-old male who does not have a good relationship with my parents (mother is 59 and father is 60) and I am suffering from mental health issues. I have an appointment at ER tomorrow, but I have not told my parents.''} The same user in an earlier post wrote -- \textit{``I am 19 year old male, a 19 year old female friend of mine is interested in dating me because I am less masculine as compared to her previous boyfriends.''} 
Here, by combining these posts, a third party can study the user's age, gender, and health, alongside garnering information about the user's parents and potential partner, along with the reason why she wanted to date the user.

With the above in-mind, this paper characterizes the multifaceted nature of self-disclosure on Reddit, across 11 distinct categories related to identity and sensitive information~\cite{dou_reducing_2024}. These categories encompass age, gender, religion, ethnicity, sexual orientation, and more. 
To study how these patterns vary across communities, we study a diverse pool of Reddit users from the top-10 (by number of users) subreddits. We then use the outcomes to design a tool that can alert users to (potentially unknown) disclosures in their social media posts. Our contributions are:

\begin{itemize}

    \item We design a novel classifier to detect the presence of 11 different types of self-disclosures in a piece of text. We make our classifier open-source to assist other researchers. (\S\ref{sec:sd_types_and_classification})
    
    \item We show that at least 50\% of users self-disclose in at least 10\% of their total posts. Moreover, 50\% of disclosing posts have more than one type of self-disclosure, revealing prior works miss significant volumes of information~\cite{zani_motivating_2022,masur_impact_2023}. We build on this to identify the social norms of self-disclosure by highlighting pairs more likely to co-occur in a post, such as \age and \gender, or \gender and \relationship. (\S\ref{sec:sd_types_co_occurance})

    \item We show that users' engagement varies significantly with the type of self-disclosure in the posts. For instance, \sexualOrientation gets 2.6x more comments than posts without any disclosure. Moreover, users who have received interactions from self-disclosure-specific community members (such as the LGBT subreddit) are more likely to disclose in the future than those who have not. (\S\ref{sec:engagement_comb})

    \item We embed our work in a browser tool that can automatically alert users to inadvertent online self-disclosure. (\S\ref{sec:tool})

\end{itemize}

\section{Background}
\label{sec:related_work}

\pb{Self-Disclosure Types and Detection.} Most of the prior literature on self-disclosure focuses on one type of self-disclosure or focused user groups and communities. For example, disclosure about mental health~\cite{choudhury_mental_2014,balani_detecting_2015} and subreddits related to health support~\cite{balani_detecting_2015}, empathy and intimacy~\cite{reuel_measuring_2022}. Some studies have looked at more than one type of self-disclosure. However, such studies are based on the discourse related to specific events such as the COVID-19 pandemic~\cite{blose_privacy_2020}. Qualitative methods such as manual coding~\cite{chen_i_2024}, surveys, and interviews~\cite{zani_motivating_2022} have been used to detect self-disclosure. 
Other researchers have used quantitative methods, consisting of supervised learning~\cite{balani_detecting_2015,wang_modeling_2016}, unsupervised learning~\cite{blose_privacy_2020}, and large language models (LLMs)~\cite{dou_reducing_2024} for self-disclosure detection.
A key contrast between our work and the above is that we focus on identifying multiple types of self-disclosure.

\pb{Self-Disclosure Characterization} 
Online self-disclosure characterization is a multidimensional area~\cite{zillich_norms_2019,bertaglia_influencer_2024}.
Several studies focus on self-disclosure as means of users' support~\cite{choudhury_mental_2014,lee_designing_2020,gaur_let_2018}, other have explored linguistic characteristics within self-disclosure~\cite{gaur_let_2018}, and its impact on privacy~\cite{dinev_privacy_2006,yun_chronological_2019,bazarova_self-disclosure_2014}. One particular direction focuses on investigating social and communication norms among users, which can lead to nuances in self-disclosure behavior~\cite{gilroy_digital_2021,dietz-uhler_formation_2005}. For example, establishment of self-disclosure norms in mental health discussions as part of communication reinforces self-disclosure~\cite{dietz-uhler_formation_2005}. However, differences can be observed within demographics; for example, younger people are more open about their sexuality~\cite{gilroy_digital_2021}. Users' gender may play an important role in their disclosure and reaction to the disclosure of others within the blogging community~\cite{jang_non-directed_2011}.

Our work is distinct in that it focuses on characterizing \emph{multiple} types of self-disclosure of a \emph{general} set of users, spanning various communities. Thus, our insights generalize to a wide population of Reddit users, and captures a general scale of disclosure.

\pb{Self-Disclosure and Privacy.} Another key line of work focuses on building tools for end-users to help control their disclosure~\cite{dou_reducing_2024,guarino_automatic_2022}.
A recent study proposed a task to help users rewrite the disclosure in their social media posts~\cite{dou_reducing_2024}. 
Similarly, Guarino et al. developed a web browser extension to help users control their disclosure based on keywords. However, it does not cover six of our identified disclosure types and relies on multiple classifiers, increasing the computational cost~\cite{guarino_automatic_2022}.
Our work provides data-driven insights and a tool to improve user privacy. Our tool covers 11 types of self-disclosure and does not require maintaining users' profiles. Additionally, our co-occurrence analysis offers insights into self-disclosures likely to occur together, which is useful in downstream research to increase the efficiency of privacy-preserving methods.

\section{Data Collection Methodology} 
\label{sec:dataset}

Our data collection aims to solicit \emph{all} posts from a Reddit user in a given time window. This ensures a holistic view of self-disclosure by a given user. \ehsan{Thus, we use the pushshift data dump of all Reddit from October 2020 to June 2021~\cite{haq_understanding_2023}, making it possible to track a user's all interactions within this time period.} We first gather a seed list of users to start our data collection. For this, we extract all users who have written a post in any of the top 10 largest subreddits\footnote{The subreddits are \textit{funny, AskReddit, gaming, aww, worldnews, todayilearned, Music, movies, science, pics}} (by community size) during two months (Jan. and Feb. 2021). We call these \emph{general users}.
Note that posts collected from these general subreddits are more likely to contain general discourse than topic-specific discourse like health and intimacy~\cite{choudhury_mental_2014,balani_detecting_2015}. 

We then gather all their posts for these general users, from October 2020 to June 2021. The raw dataset contains 16,706,119 posts from 365,385 users. We remove accounts containing any single or combination of the words \textit{`bot', `moderat', or `auto'} in their usernames to filter bot and moderator accounts~\cite{zhu2024study}. This removed 1,103 accounts and 428,275 posts.

\section{Self-Disclosure Classification} \label{sec:sd_types_and_classification}

To analyze self-disclosure at scale, it is first necessary to devise a methodology that can automatically identify the presence of disclosure in user posts. Thus, we first design a classifier.

\subsection{Defining Self-Disclosure Types}

We start by defining a taxonomy of self-disclosure types to form our supervised label set. Given our focus on privacy, we focus on self-disclosures that may reveal one's identity and information that may be used in a potential prejudice, such as health, age, gender, and sexual orientation~\cite{lee_designing_2020,wang_modeling_2016}. A recent study proposed a list of 19 self-disclosure types related to demographics and personal experiences~\cite{dou_reducing_2024}. We take inspiration from these 19 categories and a review of different self-disclosure types studied in prior literature~\cite{zani_motivating_2022}. Based on our analysis, in Table~\ref{tab:all_sd_types} (Appendix~\ref{app:sd_type_table}), we present the high-level categories of self-disclosure. These cover various aspects of one's life, including identity, relationship, work, health, group affiliations, and opinions.
For simplicity, we do not consider the group affiliations and opinion categories. This is because these categories include far more subjective information. Furthermore, it is challenging to judge if this subjective information carries sensitive insight.

Note that we consider relationship, profession/economics, and health as broader categories in this study. For instance, we consider health to be one type of self-disclosure, as our research questions do not strive to differentiate between mental and physical health at a granular scale. Similarly, finance, profession, and job-related self-disclosure are grouped as ``Job.'' These considerations help us improve our classifier's performance and answer our research questions more effectively.
Our final list of self-disclosure types is: \textbf{Age, Education, Ethnicity, Gender, Health, Job, Location, Physical Appearance, Relationship, Religion, and Sexual Orientation.} We acknowledge that our selected self-disclosure types do not form an exhaustive list. However, this list matches commonly studied self-disclosure types~\cite{zani_motivating_2022} and covers the GDPR definition of personal data, except affiliations, opinions, and genetic data~\cite{noauthor_what_2023}.

\subsection{Self-Disclosure Training Data Annotation} \label{sec:classifciation}

It is next necessary to label a training dataset, for which we take a two-step approach. To optimize the training sample, we first use ChatGPT to extract more relevant posts for manual inspection. Then, two researchers label the posts to create the training set.

\pb{Identification of Self-Disclosure.} 
There are two commonly used approaches for self-disclosure identification: \one A binary coding (\textit{yes, no}) representing the presence or absence of self-disclosure~\cite{kou_what_2018};
and
\two Identifying self-disclosure as an ordinal variable that quantifies the sensitivity of the information provided by a user~\cite{wang_modeling_2016}. Usually, this is in the form of \textit{low, medium, and high}. There is, however, no fixed criteria for using either of the approaches.
We follow the first approach and use a binary variable to indicate the presence or absence of a particular type of self-disclosure. This is because we are interested in the overall presence of self-disclosure instead of a fine-grained analysis within each self-disclosure type. We also note that an ordinal approach will require a more extensive data annotation exercise and will likely result in reduced performance.

\pb{Data Annotation.} We use a two-step data annotation exercise, employing ChatGPT (3.5) followed by manual annotation. The use of ChatGPT is to improve the sample selection, as only the samples with positive outcomes from ChatGPT are subsequently manually analyzed. We randomly select 2000 samples from the dataset for ChatGPT annotation. We then use the standard ChatGPT guidelines on the chain of thought~\cite{wei_chain--thought_2022} along with a role assignment~\cite{imran_analyzing_2023} to design our prompt (prompt in Appendix~\ref{app:chatgpt_annotation}). The GPT annotations are performed in a single session, and 1,764 posts are selected as containing at least one type of disclosure. 
Two researchers then perform manual validation of these annotations. After a briefing session, the two researchers are assigned a pool of 110 posts (taken from the positive ChatGPT annotations), consisting of 10 posts randomly selected for each of the 11 self-disclosure types. 
The results from the two researchers are compared and disagreements are discussed. Inter-rater reliability (IRR) is calculated, and the Cohen's Kappa score is in the range of 0.51 to 0.80, (median 0.70 and mean 0.70) for all labels. In addition, 19 posts are identified to be completely incorrectly classified as self-disclosing by ChatGPT, with another 30\% posts missing at least one label identified in manual inspection (where both raters have agreements). After the disagreements are discussed between the researchers, another round of manual annotations is followed on a sample of 250 posts, followed by the IRR measurements for round 2. Here, the Cohen's Kappa score for each self-disclosure type is in the range of 0.79 to 0.94 (median 0.85, and mean 0.82); showing a high agreement between the two annotators~\cite{hadi_mogavi_student_2021}. After the second debriefing round, following prior work~\cite{mcdonald_reliability_2019}, one researcher continues with the rest of the manual annotations. In total, 1,329 posts are labeled as self-disclosing. One post can have more than one type of self-disclosure. We use this final manually annotated dataset for training the classifier.

\subsection{Classifier Training}
Fine-tuning of pre-trained large models has shown positive results in downstream research tasks like ours. 
Thus, we use the labeled data to fine-tune pre-trained BERT and Roberta models~\cite{turc_well-read_2019,liu_roberta_2019}. We define this as a multi-label classification task --- the input is a post, and the output is the probability of each type of self-disclosure it contains (if any). We stratify the labeled data into 70-15-15\% for training, testing, and validation.

Roberta performs better than BERT, with an average weighted F1 of 0.83. Table~\ref{tab:classification_details} (Appendix~\ref{sec:classifier_performance}) reports the F1 scores within each self-disclosure type, along with precision and recall scores. We achieve the best F1 scores for Ethnicity (0.95), Religion (0.94), and Age (0.89). All self-disclosure except \gender and \job have >0.8 F1 score. Note that we consider a positive class if the sigmoid function has a value of 0.5 or higher. Any change in this limit can affect the data size; however, 0.5 gives a moderate approach and is commonly used in classification tasks. We use this model to predict the self-disclosure labels of every post in our dataset. We predict labels for the text and title of posts separately due to difference in the length and writing styles. For each post, we then take the union of the self-disclosure labels identified in both the text and title. Our classifier is open source for use by the community.\footnote{\url{https://huggingface.co/euhaq/self_disclosure}}

\begin{figure}[t]
    \centering
    \begin{subfigure}[t]{.2\textwidth}
        \centering
        \includegraphics[width=.85\textwidth]{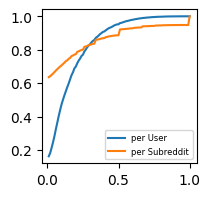} 
        \caption{Ratio of self disclosure posts across all posts}
        \label{fig:ratio_sd_posts}
    \end{subfigure}
    \quad
    \begin{subfigure}[t]{.2\textwidth}
        \centering
        \includegraphics[width=.85\textwidth]{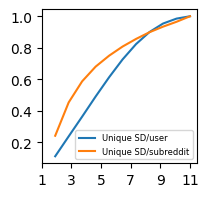}
        \caption{Unique self-disclosures per user and per subreddit}
        \label{fig:number_unique_sd_users}
    \end{subfigure}
    \quad
    \begin{subfigure}[t]{.23\textwidth}
        \centering
        \includegraphics[width=.8\textwidth]{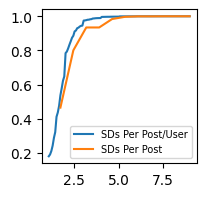}
        \caption{Unique self-disclosure types per post and per user}
        \label{fig:sd_per_post_user}
    \end{subfigure}    
    \label{fig:rq_1}
    \caption{Cumulative distributions of self-disclosure.\vspace{-10pt}}
\end{figure}

\section{Quantifying Self-Disclosure} \label{sec:sd_types_co_occurance}

Our core goal is to understand the self-disclosure norms of Reddit users.
Thus, exploiting our labels, we now measure the scale of self-disclosures to understand what sort of (privacy-sensitive) information is disclosed.

\subsection{Quantifying the Scale of Self-Disclosure} 
First, we evaluate the count self-disclosures across all users and posts, to quantify potential privacy-compromising exposure.

\pb{Number of Self-Disclosures Per User.}
For each user, we inspect the ratio of posts that contain at least one type of self-disclosure vs.\ the total number of posts from that user. Figure~\ref{fig:ratio_sd_posts} shows the distribution of this ratio for all users in our dataset. The mean of the ratio is 0.15 (median 0.10, 75th percentile 0.23). 
Half of the users in our dataset self-disclose in at least 10\% of their posts, and 25\% of users have self-disclosure in at least 23\% of their posts. This confirms that a significant number of users are involved in self-disclosure.
We emphasize that our dataset is based on users from general purpose subreddits, and we do not focus on specific subreddits that are more likely to have self-disclosure. Thus, this is surprisingly high. 

\pb{Number of Self-Disclosure Types Per User.} 
Whereas the above may raise privacy concerns, it is less problematic if a user discloses the same information repeatedly (rather than many instances of new information). Thus, we inspect the number of \emph{unique} self-disclosure types per user.
For example, if a user shares \age in four posts and \gender in two posts, this user still only has two unique self-disclosure types (\age and \gender). 

Figure~\ref{fig:number_unique_sd_users} plots the number of unique self-disclosure types per user. We see that users share an average of four ($\sigma =  2$) different types of self-disclosure across their timelines. In fact, 50\% of users have at least five types of self-disclosures in their timelines. Thus, it is clear that most users exhibit a diverse range of disclosures. 
In many cases, we find that these multiple disclosures are embedded within individual posts.
To quantify this, Figure~\ref{fig:sd_per_post_user} (orange line) plots the distribution of the number of unique disclosures per post, and the blue line shows the distribution of the per-user average of this measure.
We find that, on average, posts contain close to 2 ($\mu =1.8$, $\sigma = 2$) self-disclosure types.
This confirms that users expose substantial and diverse information. We posit that it is possible for third parties to easily combine such information. This naturally increases users' susceptibility to malicious actors in online spaces.

\begin{figure}[t]
    \centering
        \centering
        \includegraphics[width=.35\textwidth]{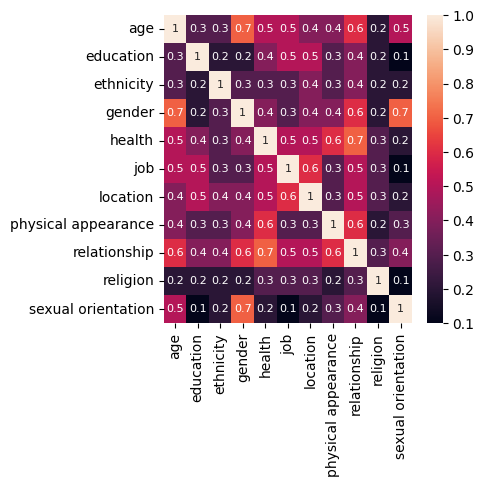}
        \caption{Correlation between different self-disclosure types that occur commonly per user. Most users who share Age also share Gender and Relationship-related self-disclosures.}
        \label{fig:sd_types_correlation_user}
\end{figure}


\begin{table*}[t]
\tiny
\caption{The likelihood of self-disclosures to co-occur in the same post. Each column and row represent model and confounding factors, respectively. $***$ indicates statistical significance. The top value displays the $\beta$ estimate and the bottom shows the standard error. Blue and Red colors indicate the most positive and negative estimates, respectively.}
\begin{tabular}{|l|l|l|l|l|l|l|l|l|l|l|l|}
\toprule
                              & \textbf{Age}                     & \textbf{Education}               & \textbf{Ethnicity}               & \textbf{Gender}                  & \textbf{Health}                  & \textbf{Job}                     & \textbf{Location}                & \textbf{Phy. Appearance}         & \textbf{Relationship}            & \textbf{Religion}                & \textbf{Sexual Orientation}      \\
\midrule
\textbf{age}                  &                                  & {\color{blue}  0.013***}  & -0.005***                        & {\color{blue}  0.271***}  & {\color{blue}  0.012***}  & {\color{blue}  0.006***}  & 0.002*                           & -0.020***                        & 0.119***                         & -0.008***                        & 0.002***                         \\
                              &                                  & 0.001                           & 0                                & 0.001                           & 0.001                           & 0.001                           & 0.001                           & 0.001                           & 0.001                           & 0.001                           & 0                                \\
\hline
\textbf{education}            & 0.017***                         &                                  & -0.014***                        & -0.053***                        & -0.168***                        & -0.136***                        & -0.112***                        & -0.057***                        & -0.103***                        & -0.045***                        & -0.023***                        \\
                              & 0.001                           &                                  & 0.001                           & 0.001                           & 0.001                           & 0.001                           & 0.001                           & 0.001                           & 0.001                           & 0.001                           & 0.001                           \\
                              
                              \hline

\textbf{ethnicity}            & {\color{red}  -0.019***} & -0.038***                        &                                  & 0.069***                         & -0.166***                        & -0.198***                        & {\color{blue}  0.009***}  & -0.005***                        & -0.087***                        & -0.029***                        & -0.003**                         \\
                              & 0.002                            & 0.002                            &                                  & 0.001                           & 0.002                            & 0.002                            & 0.002                            & 0.001                           & 0.002                            & 0.001                           & 0.001                           \\
                              
                              \hline
\textbf{gender}               & {\color{blue}  0.364***}  & -0.056***                        & {\color{blue}  0.025***}  &                                  & -0.134***                        & -0.141***                        & -0.069***                        & {\color{blue}  0.060***}  & {\color{blue}  0.141***}  & -0.023***                        & {\color{blue}  0.067***}  \\
                              & 0.001                           & 0.001                           & 0.001                           &                                  & 0.001                           & 0.001                           & 0.001                           & 0.001                           & 0.001                           & 0.001                           & 0.001                           \\
                              
                              \hline
\textbf{health}               & 0.007***                         &   -0.080*** & -0.028***                        & -0.061***                        &                                  & -0.307***                        & {\color{red}  -0.208***} & -0.002***                        & -0.147***                        & -0.063***                        & -0.036***                        \\
                              & 0.001                           & 0.001                           & 0                                & 0.001                           &                                  & 0.001                           & 0.001                           & 0.001                           & 0.001                           & 0                                & 0                                \\
                              
                              \hline
\textbf{job}                  & 0.003***                         & -0.063***                        & {\color{red}  -0.032***} & {\color{red}  -0.062***} & -0.294***                        &                                  & -0.188***                        & -0.075***                        & {\color{red}  -0.198***} & {\color{red}  -0.070***} & {\color{red}  -0.037***} \\
                              & 0.001                           & 0.001                           & 0                                & 0.001                           & 0.001                           &                                  & 0.001                           & 0.001                           & 0.001                           & 0                                & 0                                \\
                              
                              \hline
\textbf{location}             & 0.001*                           & -0.058***                        & 0.002***                         & -0.034***                        & -0.223***                        & -0.211***                        &                                  & -0.044***                        & -0.129***                        & -0.043***                        & -0.018***                        \\
                              & 0.001                           & 0.001                           & 0                                & 0.001                           & 0.001                           & 0.001                           &                                  & 0.001                           & 0.001                           & 0                                & 0                                \\
                              
                              \hline
\textbf{physical\_appearance} & -0.034***                        & -0.074***                        & -0.002***                        & 0.074***                         & -0.005***                        & -0.211***                        & -0.110***                        &                                  & -0.105***                        & -0.042***                        & -0.008***                        \\
                              & 0.001                           & 0.001                           & 0.001                           & 0.001                           & 0.001                           & 0.001                           & 0.001                           &                                  & 0.001                           & 0.001                           & 0.001                           \\
                              
                              \hline
\textbf{relationship}         & 0.069***                         & -0.047***                        & -0.014***                        & 0.062***                         & -0.140***                        & -0.197***                        & -0.115***                        & -0.037***                        &                                  & -0.027***                        & 0.002***                         \\
                              & 0.001                           & 0.001                           & 0                                & 0.001                           & 0.001                           & 0.001                           & 0.001                           & 0.001                           &                                  & 0                                & 0                                \\
                              
                              \hline
\textbf{religion}             & -0.024***                        & {\color{red}  -0.107***} & -0.024***                        & -0.051***                        & {\color{red}  -0.307***} & {\color{red}  -0.357***} & -0.198***                        & {\color{red}  -0.076***} & -0.138***                        &                                  & -0.032***                        \\
                              & 0.002                            & 0.001                           & 0.001                           & 0.001                           & 0.002                            & 0.002                            & 0.002                            & 0.001                           & 0.002                            &                                  & 0.001                           \\
                              
                              \hline
\textbf{sexual\_orientation}  & 0.008***                         & -0.065***                        & -0.003**                         & 0.179***                         & -0.210***                        & -0.227***                        & -0.095***                        & -0.017***                        & 0.012***                         & -0.038***                        &                                  \\
                              & 0.002                            & 0.001                           & 0.001                           & 0.001                           & 0.002                            & 0.002                            & 0.002                            & 0.001                           & 0.002                            & 0.001                           &    \\
                              \bottomrule
\end{tabular}
\label{tab:co_occur_regression}
\end{table*}

\begin{figure}[t]
    \centering
    \includegraphics[width=0.3\textwidth]{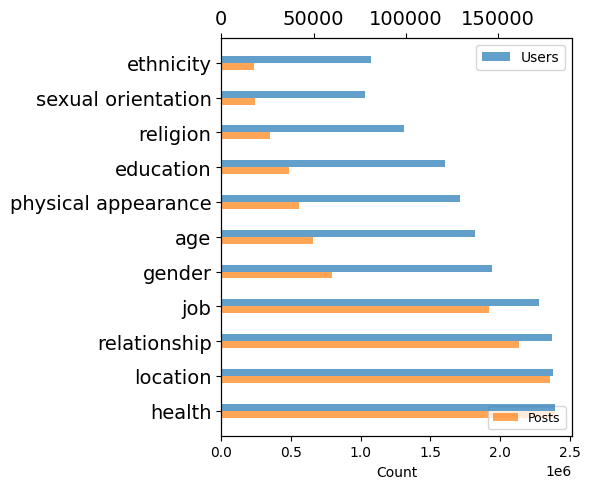}
    \caption{Number of users (top x-axis) and posts (bottom x-axis) for each self-disclosure type.\vspace{-9pt}}
    \label{fig:sd_frequency_overall}
\end{figure}

\subsection{Quantifying Types of Self-Disclosure} \label{sec:sd_correlation}

The above has identified that a large number of self-disclosures are exposed by users. We proceed to study the specific categories of information disclosed and which are co-located within posts.

\pb{Frequency of Self-Disclosure Types.}
Figure~\ref{fig:sd_frequency_overall} 
presents a histogram of the number of posts and users who disclose each type of information. Health, location, and relationship are the most commonly shared. Location is particularly concerning, as this can expose users to physical risks. Equally, health information tends to be particularly sensitive. That said, support-seeking for users with medical issues is commonplace, and some may argue the benefits outweigh the risks in certain scenarios~\cite{silveira_fraga_online_2018,zou_self-disclosure_2024}.
Information about ethnicity, sexual orientation, and religion is the least shared. these disclosures can also be used to harm users, \eg sharing such information can be used for catfishing and cyberbullying~\cite{lauckner_catfishing_2019}.

\pb{Co-occurrence of Self-Disclosures.}  
Next, we inspect which combinations of disclosure types are exposed by each user. 
We wish to understand if certain combinations of information disclosure are common (\eg revealing \emph{both} age and gender).
For this, we first compute the sum of all types of self-disclosure shared by each user. This generates a matrix of $n x 11$, where $n$ is the total number of users (recall, we cover 11 types of self-disclosure). 
We then calculate the Pearson correlation for each pair of self-disclosure types.
Figure~\ref{fig:sd_types_correlation_user} shows the correlations on a per-user level. Each cell in the figure shows the correlation between a pair of self-disclosure types --- a higher value indicates that the self-disclosure types from the pair have been shared together by more users. We do see certain pairs commonly occurring, suggesting that certain self-disclosure norms have emerged within Reddit.
Importantly, we emphasize that this disclosure behavior is not limited to support-seeking subreddits, but is instead a \textit{general} self-disclosure norm.

\pb{Modeling Self-Disclosure Relationships.}
To systematize the above analysis, we model the precise relationships using 11 different fixed effect regression models, as follows:

\begin{equation}
    s_j =  \sum_i^{11} \beta_is_i + U + T\quad  i,j \in [1,11], i\neq j
\end{equation}
where, $s_j$ is the dependent variable for the $j^{th}$ self-disclosure, and $B_i$ is the estimate (impact) of the remaining self-disclosure types. $U$ and $T$ are the fixed effects for users and time. We develop 11 regression models, one for each self-disclosure type, with the goal of identifying potential patterns of risky co-disclosure. 

Table~\ref{tab:co_occur_regression} summarizes the results, where
each column reports a regression model for a self-disclosure type mentioned in the column header. The rows of the table report the regression estimate of the remaining self-disclosure types. Additionally, each estimate is accompanied by the standard error along with p-values indicated by  $*$. For each model, we highlight the most positive estimate with blue color and the most negative one with red color. 

Confirming our previous results in Figure~\ref{fig:sd_types_correlation_user}, we observe both positive and negative correlations in self-disclosures, reflecting interesting trends.
For instance, if a post adds a self-disclosure about \gender the likelihood of \age disclosure increases by 0.364x. However, disclosure about \ethnicity will increase the likelihood of disclosing \age by just 0.02x. 
Similarly, mentioning \gender increases the likelihood of mentioning a \relationship by 0.14x, while \job reduces the likelihood of mentioning \relationship by -.198x. 
Overall, we find that \age and \gender have the largest effect sizes for most of the regression models. This implies that users most often disclose their age and gender together on Reddit. Such combinations may have benefits for support but also pose risks associated with mental health to certain demographics\eg youth during gender transitions~\cite{haimson_disclosure_2015}.
This combination is also visible in Figure~\ref{fig:sd_types_correlation_user}, which shows a 0.74 correlation between the presence of \age and \gender disclosure. Although this is arguably privacy-sensitive, it does reveal a common norm, whereby users are expected to express such information (\eg ``[30 F]'') as part of daily discussion~\cite{chen_i_2024}.

At the opposite end of the spectrum, \religion and \job generally have the least effect size for most regression models. This suggests these are least likely to co-occur with other self-disclosure types in the same post. For instance, the likelihood of \health disclosure is reduced by 0.357x with \religion.
\religion and \job generally have a negative impact on most of self-disclosure types to co-occur with them. For instance, the likelihood of \health disclosure is reduced by 0.357x with \religion. 
The positive correlation of \gender with \physical and \sexualOrientation shows that the latter two are accompanied mostly by the former. 
We argue that topics related to physical appearance, such as body shaming together with gender, may lead towards negative outcomes such as body shaming and psychological abuse~\cite{mcmahon_body_2022,corradini_dark_2023}, or hate speech concerning the sexuality of the users~\cite{crowley_expressive_2014,lingiardi_mapping_2020}.

\section{Self-Disclosure Engagement}
\label{sec:engagement_comb}

\begin{figure}[t]
    \centering
    \begin{subfigure}{.38\textwidth}
        \centering
        \includegraphics[width=.7\textwidth]{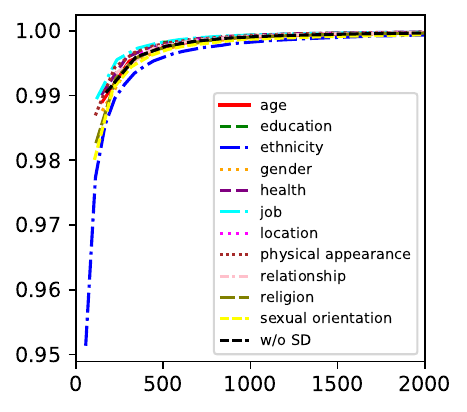}
        \caption{Number of Comments}
        \label{fig:comments}
    \end{subfigure}
    \quad

    \begin{subfigure}{.38\textwidth}
        \centering
        \includegraphics[width=.7\textwidth]{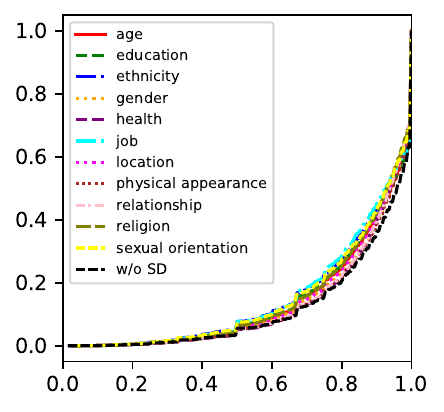}
        \caption{Upvote Ratio}
        \label{fig:upvote}
    \end{subfigure}
    \caption{Cumulative distributions of per-user mean engagement values, per disclosure type. 
    }
    \label{fig:all_engagement_cdf}
\end{figure}

\begin{figure}[t]
    \centering
    \begin{subfigure}{.5\textwidth}
        \centering
        \includegraphics[width=.85\linewidth]{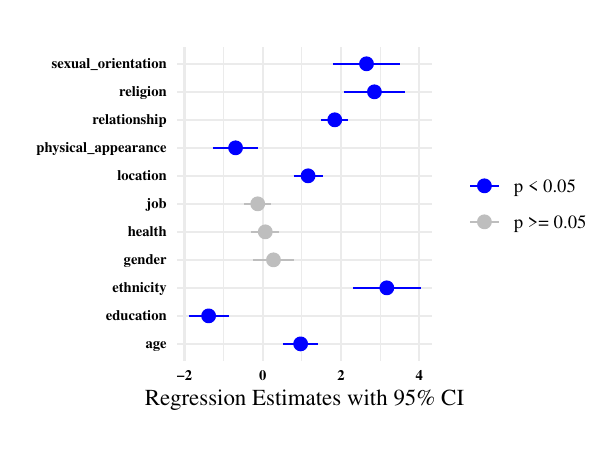}
        \caption{Effect on Number of Comments}
        \label{fig:engagement_num_comments}
    \end{subfigure}
    \quad
    \begin{subfigure}{.48\textwidth}
        \centering
        \includegraphics[width=.85\linewidth]{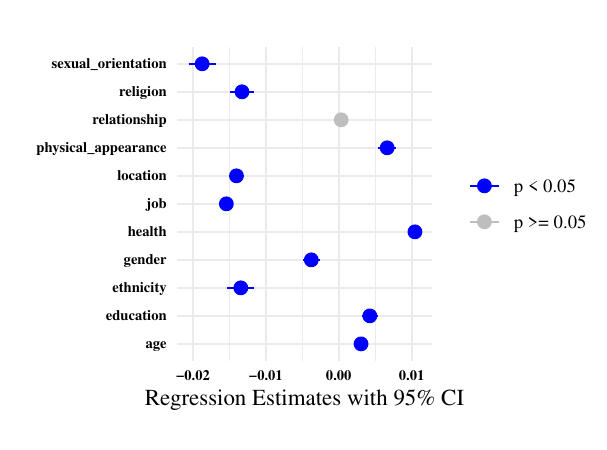}
        \caption{Effect on Upvote Ratio}
        \label{fig:engagement_upvotes_ratio}
    \end{subfigure}
    \caption{Regression results for engagement. Each panel is a different regression model. The y-axis and x-axis show confounding factors and corresponding estimate, respectively.
    }
    \label{fig:all_engagement_regression}
\end{figure}

Next, we explore user engagement with self-disclosing posts, focusing on how interactions in self-disclosure communities (\eg, /r/AMA) may encourage others to disclose, potentially creating privacy-compromising attack vectors.

\subsection{Quantifying Engagement with Self-Disclosure} \label{subsec:engagement_diff}

From a privacy perspective, higher engagement reflects potentially increased exposure to users' information. It may also create a feedback loop that influences what users post in the future~\cite{haq_short_2022}.
We therefore start by measuring the difference in engagement levels across self-disclosing posts.

Figure~\ref{fig:all_engagement_cdf} presents the distribution of two engagement metrics: \one number of comments 
and \two the ratio of upvotes to downvotes, across all posts containing self-disclosure. 
A non-parametric Kruskal-Wallis test confirms significant differences across the distribution of each disclosure type (test statistics are in Table~\ref{tab:eng_post_hoc_num_comments} and \ref{tab:eng_post_hoc_upvote_ratio} in Appendix). 
For instance, posts with \sexualOrientation have more comments ($\mu = 16.4$) than posts with \job disclosure ($\mu = 11.6$). 

To systematically analyze these differences, we turn to regression analysis with users and time-fixed effects. For each engagement metric (number of comments and upvote ratio) and self-disclosure type, we design a separate engagement prediction task based on whether the posts contain a particular self-disclosure or not (not including other types of self-disclosure). The posts without self-disclosure from all users remain identical across the models, hence providing a common baseline across models to compare the results with non-disclosing posts and within different types.

Figure~\ref{fig:engagement_num_comments} and~\ref{fig:engagement_upvotes_ratio} show the regression results for number of comments and upvote ratio, respectively. The Y-axis shows the corresponding self-disclosure used as a confounding factor. The X-axis shows the regression estimate with a 95\% confidence interval. The values in the blue color are statistically significant ($p<0.05$), and the grey color shows non-significance.

We see that the presence of self-disclosure does have a significant effect on both the number of comments and the upvote ratio. Interestingly, the effect is not similar across all self-disclosure types, and there is also a disparity in each engagement metric for a given self-disclosure. For instance, the presence of \sexualOrientation disclosure increases the number of comments (2.65x); however, the same has a negative effect on the upvote ratio($\approx-.02x$). 

To further see the difference between heterosexual and potentially more vulnerable non-hetero, we do a keyword filtering of the posts (with sexual orientation disclosure) containing the words (\textit{gay, lesbian, bisexual, and straight}), and use a non-parametric Kruskal-Wallis test ($\chi^2 = 251.2, df = 3, p < 0.001 $), followed by Dunn's test, to see the engagement metric distributions difference across each keyword's post. We observe that posts containing words `gay' receive more comments (almost double) ($\mu = 102$) and lower upvote ($\mu = .80$) than other non-hetero (\eg upvote ratio for `lesbian'  $\mu = 0.83$).
However, it receives fewer comments than the posts containing `straight' keywords ($\mu = 119$).
This contrast shows that, although \sexualOrientation increases engagement in terms of comment count, the engagement quality is less positive, on average. 
We posit that such engagement metrics may influence users' future self-disclosure likelihood, as prior Reddit studies find engagement do affect topic choice \cite{haq_short_2022}. Moreover, a lower upvote ratio with higher comments may signify negative discussions and discontent with the original poster~\cite{risch_top_2020}, potentially leading to negative experiences, especially in disclosures like \sexualOrientation and \ethnicity.

\begin{figure}[t]
    \centering
    \includegraphics[width=0.48\textwidth]{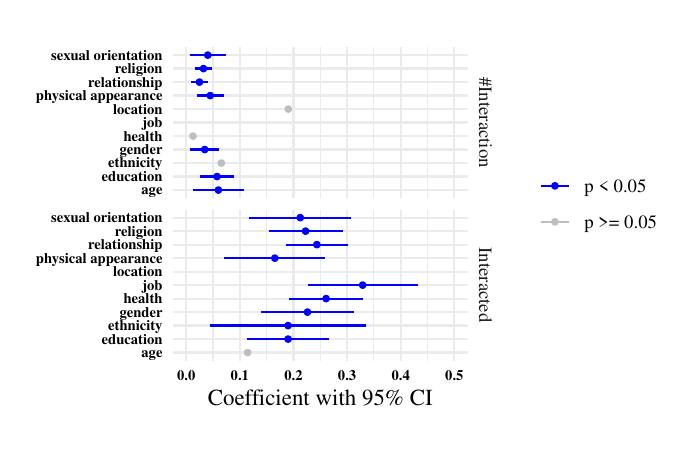}
    \caption{
    Self-disclosure types on the y-axis represent one regression model with two independent variables (\#interactions and Interacted). The x-axis shows the $\beta$ estimates.}
    \label{fig:regression_analysis_m1}
\end{figure}

\subsection{Quantifying Impact of Self-Disclosure Communities} \label{subsec:engagement_effect}

There are various Reddit communities directly related to specific forms of self-disclosure, \eg \texttt{r/aznidentity} is related to discussions on ethnicity, \texttt{AskDocs} has health-related discussions, and \texttt{r/AskMen} has gender-related discussions. 
These pertain to \ethnicity, \health, and \gender, respectively. We hypothesize that engaging with members of such communities may increase one's own likelihood of disclosing, even to other communities. This may create attack surface where malicious actors purposefully post (fake) self-disclosure to encourage others to share. We next explore the potential presence of such behaviors, quantifying the impact of receiving comments that contain self-disclosure.

\pb{Self-Disclosure related Subreddits.} 
For the above analysis, we first obtain a set of subreddits dedicated to self-disclosure. The names of the subreddits indicate their association with a focused topic~\cite{adelani_estimating_2020}. We extract the top 50 largest subreddits, in terms of their number of posts classified as containing each type of self disclosures. As some subreddits are associated with multiple self-disclosure types, we end up with 250 unique subreddits out of 550 extracted subreddits. 
We then manually review the name of each subreddit and annotate whether it is associated with one of our self-disclosure types.
Some examples are shown in Table~\ref{tab:sd_association} in Appendix~\ref{app:sd_association}.

\pb{Regression Task.} We next ask \one what is the effect of a user receiving an interaction from a users in a self-disclosure related subreddit compared to not having received an interaction? 
and \two If a user receives an interaction, what is the impact on the number of such interactions? We model this as a regression task to predict whether the user will have a self-disclosure in future: 

\[ Y_{st} = \alpha + \beta NI_{t-1} + U + T \]
$Y_{st}$ is the number of self-disclosures by a user in a period t (1 week). $\beta$ shows the coefficient for number of interactions($N_{t-1}$) at time $t-1$ with the users from selected subreddit communities (SD-specific or general subreddits). $I = [0,1]$ whether a user received an interaction $I = 1$ or not $I = 0$. 
Note, we consider interaction to be any action initiated by users from a selected community. Thus, interaction occurs when a user from a selected subreddit writes a comment (at any level of the post) on a post by our selected users.
Finally, $U$ represent all users, and $T$ represents time (in weekly brackets) to control any user and time-dependent fixed-effect. In total, we run 11 regression models. Each model is specific to the self-disclosure-specific subreddits labeled with that type of disclosure. Note, we consider any self-disclosure as a positive instance of self-disclosure and do not differentiate within different types.

\pb{Results.} Following these steps, we run our fixed effect regression model while controlling heteroskedasticity~\cite{gujarati_basic_2009}. Figure~\ref{fig:regression_analysis_m1} plots the regression model results. The figure consists of two panels, one for each of the independent variables.
The lower panel shows the binary impact of the user receiving an interaction or not, whereas the upper panel shows the impact of the number of interactions. The x-axis shows the $\beta$ estimates for variables, and the y-axis refers to the regression model depending on the self-disclosure for which the specific subreddits are used. The blue color shows the results that are statistically significant (with $p-values$ being lower than $0.05$). The gray color shows the corresponding estimates are statistically not significant. The missing values are also statistically not significant and the values are less than zero, so they are not shown due to the figure's x-axis scale.  

Confirming our hypothesis, we observe a positive effect ($\beta$ estimates) for future self-disclosure, \ie users are more likely to share a self-disclosing posts if they receive an interaction from a user who has previously posted in self-disclosure-specific subreddits.
Distinct effect sizes for the \textit{Interacted} variable are observed within each model. 
The most significant effect is observed for \job, with the odds of future self-disclosure being 0.34x, followed by \health at 0.26x, and \relationship at 0.24x. Similarly, the \textit{\#interactions} shows the positive effect for each such interaction. 
Given these confounding factors, we hypothesize two scenarios where Reddit users could be exploited: \one A user may be influenced to self-disclose through repeated interactions with other users who engage in self-disclosure, potentially including bots; or \two The creation of subreddits designed to \textit{maliciously} foster a sense of community to increase the likelihood of users engaging in self-disclosure.


\section{InsightWatcher - A Browser Plugin}\label{sec:tool}
\begin{figure}[t]
    \centering
    \begin{subfigure}{.48\textwidth}
        \centering
        \includegraphics[width=\textwidth]{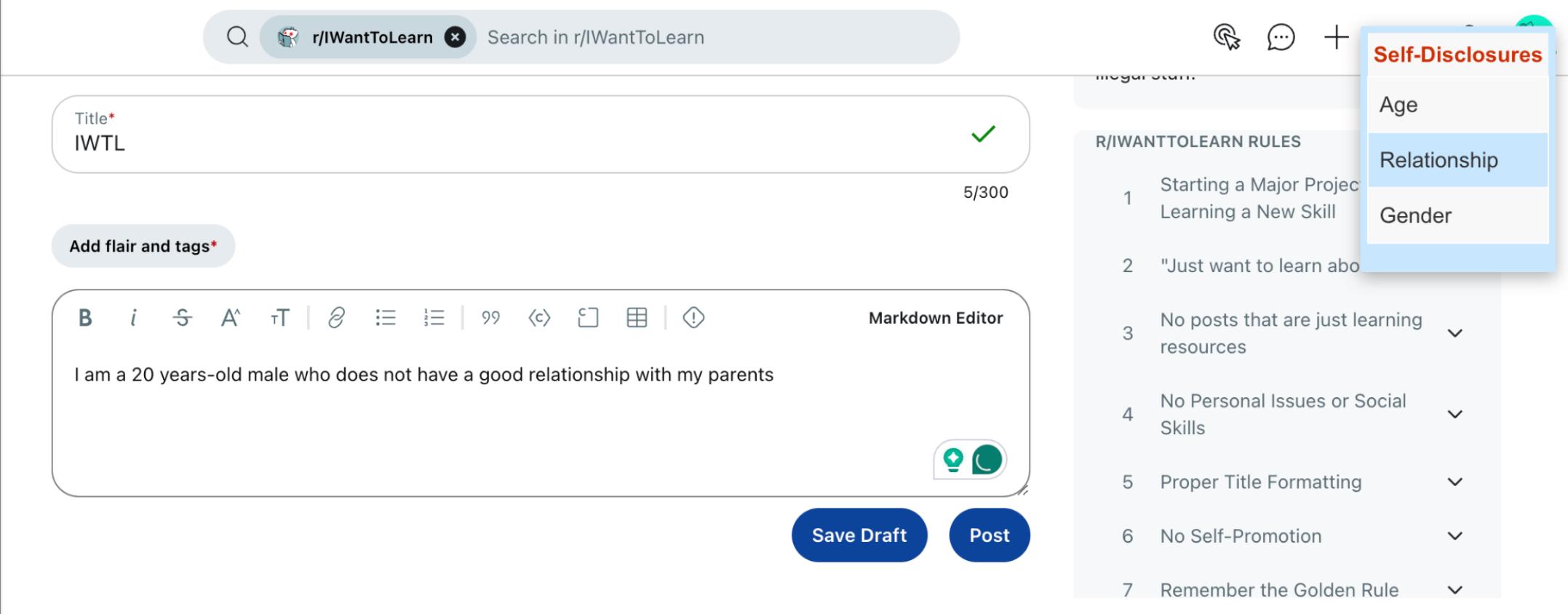}
        \caption{Active Reminder}
        \label{fig:sd_active_reminder}
    \quad
    \end{subfigure}
    \begin{subfigure}{.48\textwidth}
        \centering
        \includegraphics[width=\textwidth]{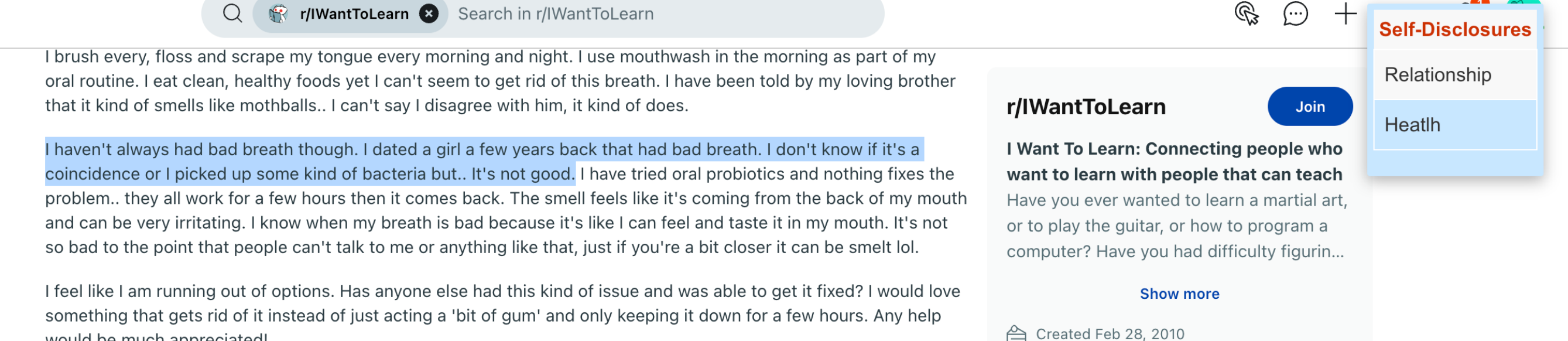}
        \caption{Passive Reminder}
        \label{fig:sd_passive_reminder}
    \end{subfigure}
    \caption{Working Example of Browser Plugin}
    \label{fig:browser_plugin}
\end{figure}

To help users control their self-disclosure, we have developed a browser plugin, \textit{InsightWatcher}. The tool automatically scans for self-disclosure, in real time, within any text box that the user loads in their browser. It works across any webpage that contains a text box (including Twitter/X, Facebook, and WhatsApp Web). Whenever a self-disclosure is identified, a small non-invasive popup is raised, notifying the user of the information they are exposing if they proceed.
The plugin achieves this using our classifier. 
It does not require any user login and does not record any inputs. 
Note, the tool also allows users to select any text on a web page, and request a list of self-disclosures within the text.
Figure~\ref{fig:sd_active_reminder} shows a screenshot of the active reminder, featuring a pop-up on the right side of the screen in real-time. In Figure~\ref{fig:sd_passive_reminder}, a user selects text from a webpage, and a pop-up displays the self-disclosures from it. The plugin will be open source and available for users install.\footnote{\url{https://github.com/ehsanulhaq1/InsightWatcher}}

\section{Discussion} \label{sec:discussion}

Our work presents the first large-scale and multi self-disclosure characterization on Reddit based on general discourse.

\pb{Privacy Control Measures.} Our work highlights that a large number of posts have self-disclosure; 50\% of users in our dataset have self-disclosure in at least 10\% of their posts. In addition, half of the users disclose more than one type of self-disclosure.  In addition to the empirical insights, there are other latent risks that lie in multiple disclosures from a user. For example, when users self-disclose, they often include information from their past and other people involved around that time. Thus, this can increase users' vulnerability to mal-actors who can combine such information to find more about a user. This behavior increases the need for tools, like our browser plugin InsightWatcher, that help users control their disclosure. \ehsan{However, Self-disclosure moderation should be context-specific, guided by ethical principles. In support-based communities, self-disclosure can enhance support, so strict moderation might lessen their positive impact. In contrast, disclosure in generic communities can be more harmful to users, thus requiring a nudge on self-disclosure.}

\pb{Co-located Self-Disclosure Types.}
Our work highlights the importance of studying multiple self-disclosures together. Users do this to add more information and context while describing their life events or sharing their thoughts. Our work highlights that pairs of self-disclosure types, such as \gender and \relationship, are likely to appear together. 
This has implications when studying such disclosures alone. The occurrence of one particular disclosure may be inspired by another, hence deviating from commonly associated characteristics with certain disclosures. The study of co-located self-disclosure can offer insights into several use-cases, such as uncovering the mechanism of identity establishment, particularly for vulnerable populations. For instance, studying \sexualOrientation together with \age and \gender and \relationship can uncover how different groups open up about their sexuality.

\pb{Disclosure about Close Contacts.}
We observe that users often disclose about people around themselves, including immediate family members, friends, and co-workers.  We find that over 88,709 posts mention at least one word from--- \textit{father, mother, sister, brother, boyfriend, girlfriend, bf, gf} in their main body. Thus, privacy risks can extend from one user to their extended social network. 
Reddit users often put such details and other characters to add more contextual information to refer to some event while sharing their experiences or seeking information. Thus, any tools designed must be flexible enough to accommodate these norms while minimizing privacy risk. We conjecture that dynamically rewriting such information, while not compromising the context could be valuable (\eg changing the specific age).

\section{Conclusion}

We have presented the first multiple-type self-disclosure characterization on Reddit. We identified 11 types of self-disclosure associated with a user's identity and demographics, such as age, gender, relationship, and job.  We first developed our open-source multi-label self-disclosure classifier for the 11 different self-disclosure types. Our characterization shows that at least half of users self-disclose in more than 10\% of their posts. We highlight that user posts are not limited to one particular type of self-disclosure. Instead, users share more than one type of information to add more context to posts. Through thematic analysis, we show that user self-disclosure reveals information about themselves and extends to their social connections, such as parents, siblings, and partners. Building on this, we have developed and deployed a browser plugin tool that can automatically notify users when they are self-disclosing. 

We note our study has several limitations, which form the basis of our future work. Most notably, our list of self-disclosure types is not exhaustive, albeit covering key identifiable information. \ehsan{Future work can extend disclosure types, such as disclosure through opinions (the types we have not included in our analysis) or increasing granularity as high, medium, and low.} \ehsan{Moreover, our work is limited to Reddit. However, a similar approach can be extended to other platforms; for example, self-disclosure-related Facebook groups or hashtag-specific discussions can be taken as a proxy of Reddit communities for Facebook and X, respectively.} Furthermore, our work is focused only on Reddit; similar studies are required on other platforms to analyze the generalization of our findings.
Finally, within our work, we consider the presence of self-disclosure but do not quantify whether the self-disclosure is high or low risk. Our future work will focus on better understanding the exact nature of the risks involved and how they can be mitigated.


\section{Acknowledgments}
This work was supported in part by the Guangzhou Science and Technology Bureau (2024A03J0684); the Guangzhou Municipal Science and Technology Project (2023A03J0011); the Guangzhou Municipal Key Laboratory on Future Networked Systems (024A03J0623), the Guangdong Provincial Key Lab of Integrated Communication, Sensing and Computation for Ubiquitous Internet of Things (No. 2023B1212010007); and by AP4L (EP/W032473/1).

\bibliographystyle{ACM-Reference-Format}
\balance
\bibliography{ref}


\appendix


\section{Ethics Statement}
This research complies with the SAGE (Self-Assessment Governance and Ethics Form for Humans and Data Research) self-check process provided by the University of Surrey, UK 
for ethics approval. No governance risks or ethical concerns falling under the higher, medium, or lower risk criteria were identified so  Ethics and Governance Application (EGA) was not required for this study. No unauthorized access or collection of private data has occurred during this research. All datasets created and collected are sourced from publicly available materials. Reddit, a platform used in this project, openly shares its content as free and open data. Since this project does not involve interactions with human subjects, there was no requirement for informed consent. We confirm compliance with the University’s Code on Good Research Practice, Ethics Policy, and all relevant professional and regulatory guidelines. \ehsan{We highlight that we do not use the data to identify any user. The analysis is performed and reported at aggregate levels, thus minimizing the privacy risks. The examples shown in the paper are chosen at random, and the text is paraphrased (e.g., line 75, page 1) so that readers cannot directly search for the same text on Reddit.}

\section{ChatGPT Annotation Prompt}\label{app:chatgpt_annotation}

\fbox{
    \begin{minipage}{.45\textwidth}
    \texttt{
            \{ "Prompt": "You are a highly talented assistant to annotate the text. You will be helping me to identify the self-disclosure in the following prompts. Self-disclosure is defined as revealing personal information to other. Your goal is to analyze the text field and see if there is any self-disclosure and add the response as yes or no in SD field, and if there is any self-disclosure related to any of the labels in the label, list them in label field. You can use more than one label if they match.",\\
            "Text": " \hl{Proud Kashmiri Pandit}",\\
            "Labels": [`age', `education', 1ethnicity', `gender', `health', `job', `location', `physical appearance', `relationship', `religion', `sexual orientation'],\\
            "Desired format": \{\\
            "SD": "",\\
            "labels":[]\}\\
                \textbf{Response} = 
                \{\\
                     "Desired format": \{ \\ 
                      "SD": "yes",\\
                    "labels":  [\hl{`ethnicity}',\hl{`religion'}]\}\\
                \}  
    }
    \end{minipage}
}


\section{Self-Disclosure Types}\label{app:sd_type_table}
Table~\ref{tab:all_sd_types} shows the disclosure types identified through the literature review in this paper. The highlighted boxes show the types that are used in this paper.

\begin{table*}[h]
\small
\caption{Self-Dislcosure Types and Categories: Highlighted color shows the types that are used in the paper.}
\begin{tabular}{|l|l|p{1.6cm}|l|p{3.5cm}|p{4.5cm}|}
\toprule
\textbf{Identity} & \textbf{ \cellcolor{yellow}Relationship} &  \textbf{Profession/ Economic} & \textbf{ \cellcolor{yellow}Health} & \textbf{Group Affiliation} & 
\textbf{Opinions/ Interests/ Feelings} \\
\midrule
   \cellcolor{yellow}Birthday/Age & Family  & \cellcolor{yellow}Job/Finance & General Health & \cellcolor{yellow}Religion & Sports \\
  \cellcolor{yellow}Ethnicity & Relations &   \cellcolor{yellow}Education & Physical Health & Politics & Politics \\
  \cellcolor{yellow}Sexual Orientation & Friendship   & & Mental Health & Community (offline vs online) &  Current-Affairs\\
 \cellcolor{yellow}Location &   &  &  &  &Religion  \\
  \cellcolor{yellow}Physical Appearance &    &  &  &  & Administration \\
  \cellcolor{yellow}Gender &  &  &  &  &  \\
 \bottomrule
\end{tabular}
\label{tab:all_sd_types}
\end{table*}

\section{Self-Disclosure Specific Communities}\label{app:sd_association}
Table~\ref{tab:sd_association} lists subreddits specific to \sexualOrientation self-disclosure. These subreddits are more likely to have disclosures related to the given type of self-disclosure. In contrast the general category of subreddit are not restricted to any self-disclosure. 
\begin{table}[H]
    \centering
        \caption{Examplars of subreddits' association with self-disclosure.}
    \begin{tabular}{|p{2.5cm}|p{.3\textwidth}|}
    \toprule
        \textbf{Self-disclosure Type} &\textbf{Subreddits} \\
        \midrule
Sexual Orientation & `lgbt', `bisexual', `askgaybros', `BisexualTeens', `LGBTeens', `actuallesbians', `gay', `asexuality', `AreTheStraightsOK', `me\_irlgbt', `SuddenlyGay', `comingout', `pansexual', `TwoXChromosomes', `AskGayMen', `gaybros'  \\
\hline
General Subreddits & `AskReddit', `memes', `cats', `Showerthoughts'\\
\bottomrule
\end{tabular}
\label{tab:sd_association}
\end{table}

\section{Classifier Performance}\label{sec:classifier_performance}

Table~\ref{tab:classification_details} shows the performance metrics of the classifier.

\begin{table}[H]
\caption{$F1$ performance for Roberta Fine-tuning.}
\begin{tabular}{l|l|l|l}
\toprule
\textbf{Self-Disclosure}     & \textbf{Precision} & \textbf{Recall} & \textbf{F1} \\
\midrule
\textbf{Age}                 & 0.84               & 0.93            & 0.89        \\
\textbf{Ethnicity}           & 1.00               & 0.90            & 0.95        \\
\textbf{Gender}              & 0.86               & 0.60            & 0.71        \\
\textbf{Education}           & 0.87               & 0.81            & 0.84        \\
\textbf{Health}              & 0.88               & 0.84            & 0.86        \\
\textbf{Job}                 & 0.89               & 0.70            & 0.78        \\
\textbf{Location}            & 0.79               & 0.88            & 0.83        \\
\textbf{Physical Appearance} & 0.81               & 0.87            & 0.84        \\
\textbf{Relationship}        & 0.84               & 0.86            & 0.85        \\
\textbf{Religion}            & 0.88               & 1.00            & 0.94        \\
\textbf{Sexual orientation}  & 0.88               & 0.79            & 0.83 \\
\bottomrule
\end{tabular}
\label{tab:classification_details}
\end{table}

\section{Engagement Statistics}
Table~\ref{tab:eng_post_hoc_num_comments} and~\ref{tab:eng_post_hoc_upvote_ratio} show the pairwise Dunn test results for the engagement parameters. 

\begin{table*}[h]
\small
\caption{Dunn's test for pairwise comparison for the number of comments across self-disclosure types. A positive value indicates that self-disclosure in the row has a higher mean value than the one in the column header. P-values are adjusted according to the Bonferroni method.  Kruskal-Walis Test = ($\chi^2  = 10484, df = 10, ***$), ($* = p <0.05, ** = p <0.01, *** = p <0.001 $) }
\begin{tabular}{p{1.6cm}|llllllp{1.5cm}llp{1.2cm}}
\toprule
\textbf{SD}                  & \textbf{education} & \textbf{ethnicity} & \textbf{gender} & \textbf{health} & \textbf{job} & \textbf{location} & \textbf{phy. appearance} & \textbf{relationship} & \textbf{religion} & \textbf{sexual orient.} \\
\midrule
\textbf{age}                 & 15.8 ***           & 16.6 ***           & 50.4 ***        & 24.8 ***        & 21.8 ***     & 72.9 ***          & 48.3 ***                     & 25.1 ***              & 20.4 ***          & 31.5 ***                    \\
\hline
\textbf{education}           &                    & 3.7 **             & 30.8 ***        & 4.0 **          & 1.6          & 48.1 ***          & 30.1 ***                     & 4.4 ***               & 5.9 ***           & 18.3 ***                    \\
\hline
\textbf{ethnicity}           &                    &                    & 20.5 ***        & -1.3            & -3.1         & 32.1 ***          & 20.7 ***                     & -0.9                  & 1.5               & 12.7 ***                    \\
\hline
\textbf{gender}              &                    &                    &                 & -36.3 ***       & -38.3 ***    & 15.2 ***          & 1.3                          & -35.4 ***             & -21.2 ***         & -4.9 ***                    \\
\hline
\textbf{health}              &                    &                    &                 &                 & -3.6 *       & 66.7 ***          & 34.2 ***                     & 0.7                   & 3.6 *             & 18.0 ***                    \\
\hline
\textbf{job}                 &                    &                    &                 &                 &              & 68.0 ***          & 36.2 ***                     & 4.2 **                & 5.6 ***           & 19.6 ***                    \\
\hline
\textbf{location}            &                    &                    &                 &                 &              &                   & -12.2 ***                    & -65.0 ***             & -34.9 ***         & -15.1 ***                   \\
\hline
\textbf{phy. appearance} &                    &                    &                 &                 &              &                   &                              & -33.5 ***             & -21.3 ***         & -5.7 ***                    \\
\hline
\textbf{relationship}        &                    &                    &                 &                 &              &                   &                              &                       & 3.2               & 17.6 ***                    \\
\hline
\textbf{religion}            &                    &                    &                 &                 &              &                   &                              &                       &                   & 12.2 ***          \\
\bottomrule
\end{tabular}
\label{tab:eng_post_hoc_num_comments}
\end{table*}

\begin{table*}[h]
\small
\caption{Dunn's test for pairwise comparison for the upvote ratio across self-disclosure types. A positive value indicates that self-disclosure in the row has a higher mean value than the one in the column header. P-values are adjusted according to the Bonferroni method.  Kruskal-Walis Test = ($\chi^2  = 4322.7, df = 10, ***$), ($* = p <0.05, ** = p <0.01, *** = p <0.001 $) }
\begin{tabular}{p{1.6cm}|llllllp{1.5cm}llp{1.2cm}}
\toprule
\textbf{SD}                  & \textbf{education} & \textbf{ethnicity} & \textbf{gender} & \textbf{health} & \textbf{job} & \textbf{location} & \textbf{phy. appearance} & \textbf{relationship} & \textbf{religion} & \textbf{sexual orient.} \\
\midrule
\textbf{age}                 & 2.2                & 24.2 ***           & 4.7 ***         & -0.1            & 30.7 ***     & 11.4 ***          & -8.9 ***                     & 9.7 ***               & 28.1 ***          & 26.1 ***                    \\
\hline
\textbf{education}           &                    & 21.6 ***           & 2.1             & -2.7            & 25.5 ***     & 7.8 ***           & -10.6 ***                    & 6.2 ***               & 24.9 ***          & 23.4 ***                    \\
\hline
\textbf{ethnicity}           &                    &                    & -21.3 ***       & -26.9 ***       & -5.2 ***     & -18.9 ***         & -30.7 ***                    & -20.0 ***             & 0.8               & 1.7                         \\
\hline
\textbf{gender}              &                    &                    &                 & -6.0 ***        & 26.8 ***     & 6.3 ***           & -13.9 ***                    & 4.5 ***               & 25.0 ***          & 23.2 ***                    \\
\hline
\textbf{health}              &                    &                    &                 &                 & 42.1 ***     & 15.9 ***          & -10.7 ***                    & 13.5 ***              & 32.2 ***          & 28.9 ***                    \\
\hline
\textbf{job}                 &                    &                    &                 &                 &              & -26.3 ***         & -40.0 ***                    & -28.4 ***             & 7.2 ***           & 7.4 ***                     \\
\hline
\textbf{location}            &                    &                    &                 &                 &              &                   & -21.6 ***                    & -2.3                  & 23.0 ***          & 21.0 ***                    \\
\hline
\textbf{physical appearance} &                    &                    &                 &                 &              &                   &                              & 20.0 ***              & 35.2 ***          & 32.4 ***                    \\
\hline
\textbf{relationship}        &                    &                    &                 &                 &              &                   &                              &                       & 24.2 ***          & 22.1 ***                    \\
\hline
\textbf{religion}            &                    &                    &                 &                 &              &                   &                              &                       &                   & 1       \\
\bottomrule
\end{tabular}
\label{tab:eng_post_hoc_upvote_ratio}
\end{table*}

\end{document}